\documentstyle[epsf]{mn}

\def\simlt{\lower.5ex\hbox{$\; \buildrel < \over \sim \;$}}
\def\simgt{\lower.5ex\hbox{$\; \buildrel > \over \sim \;$}}
\newif\ifAMStwofonts
\ifoldfss
  \ifCUPmtlplainloaded \else
    \NewTextAlphabet{textbfit} {cmbxti10} {}
    \NewTextAlphabet{textbfss} {cmssbx10} {}
    \NewMathAlphabet{mathbfit} {cmbxti10} {} % for math mode
    \NewMathAlphabet{mathbfss} {cmssbx10} {} %  "   "    "
  \fi
  \ifAMStwofonts
    \ifCUPmtlplainloaded \else
      \NewSymbolFont{upmath} {eurm10}
      \NewSymbolFont{AMSa} {msam10}
      \NewMathSymbol{\upi}     {0}{upmath}{19}
      \NewMathSymbol{\umu}     {0}{upmath}{16}
      \NewMathSymbol{\upartial}{0}{upmath}{40}
      \NewMathSymbol{\leqslant}{3}{AMSa}{36}
      \NewMathSymbol{\geqslant}{3}{AMSa}{3E}

    \fi
  \fi
\fi % End of OFSS

\ifnfssone
  \newmathalphabet{\mathit}
  \addtoversion{normal}{\mathit}{cmr}{m}{it}
  \addtoversion{bold}{\mathit}{cmr}{bx}{it}
  \newmathalphabet{\mathbfit} % math mode version of \textbfit{..}
  \addtoversion{normal}{\mathbfit}{cmr}{bx}{it}
  \addtoversion{bold}{\mathbfit}{cmr}{bx}{it}
  \newmathalphabet{\mathbfss} % math mode version of \textbfss{..}
  \addtoversion{normal}{\mathbfss}{cmss}{bx}{n}
  \addtoversion{bold}{\mathbfss}{cmss}{bx}{n}
  \ifAMStwofonts
    \ifCUPmtlplainloaded \else
      \UseAMStwoboldmath
      \makeatletter
      \new@mathgroup\upmath@group
      \define@mathgroup\mv@normal\upmath@group{eur}{m}{n}
      \define@mathgroup\mv@bold\upmath@group{eur}{b}{n}
      \edef\UPM{\hexnumber\upmath@group}
      \new@mathgroup\amsa@group
      \define@mathgroup\mv@normal\amsa@group{msa}{m}{n}
      \define@mathgroup\mv@bold\amsa@group{msa}{m}{n}
      \edef\AMSa{\hexnumber\amsa@group}
      \makeatother
      \mathchardef\upi="0\UPM19
      \mathchardef\umu="0\UPM16
      \mathchardef\upartial="0\UPM40
      \mathchardef\leqslant="3\AMSa36
      \mathchardef\geqslant="3\AMSa3E
    \fi
  \fi
\fi % End of NFSS release 1

\ifnfsstwo
  \DeclareMathAlphabet{\mathbfit}{OT1}{cmr}{bx}{it}
  \SetMathAlphabet\mathbfit{bold}{OT1}{cmr}{bx}{it}
  \DeclareMathAlphabet{\mathbfss}{OT1}{cmss}{bx}{n}
  \SetMathAlphabet\mathbfss{bold}{OT1}{cmss}{bx}{n}
  \ifAMStwofonts
    \ifCUPmtlplainloaded \else
      \DeclareSymbolFont{UPM}{U}{eur}{m}{n}
      \SetSymbolFont{UPM}{bold}{U}{eur}{b}{n}
      \DeclareSymbolFont{AMSa}{U}{msa}{m}{n}
      \DeclareMathSymbol{\upi}{0}{UPM}{"19}
      \DeclareMathSymbol{\umu}{0}{UPM}{"16}
      \DeclareMathSymbol{\upartial}{0}{UPM}{"40}
      \DeclareMathSymbol{\leqslant}{3}{AMSa}{"36}
      \DeclareMathSymbol{\geqslant}{3}{AMSa}{"3E}
    \fi
  \fi
\fi % End of NFSS release 2

\ifCUPmtlplainloaded \else
  \ifAMStwofonts \else % If no AMS fonts
    \def\upi{\pi}
    \def\umu{\mu}
    \def\upartial{\partial}
  \fi
\fi

\title{A Study of the New X-ray Transient RXTE~J2123$-$058 \\  
        during its Post-Outburst State} 

\author[R. Soria, K. Wu \& D. Galloway]
       {Roberto Soria$^{1,2}$, Kinwah Wu$^2$ and Duncan Galloway$^{2,3}$\\
   $^1$ Research School of Astronomy and Astrophysics, 
      Australian National University,     
      Private Bag, Weston Creek P.O., ACT 2611, \\ 
      \ \  Australia; roberto@mso.anu.edu.au \\ 
   $^2$ Research Centre for Theoretical Astrophysics, School of Physics, 
      University of Sydney, NSW 2006, Australia \\
   $^3$ Physics Department, University of Tasmania, GPO Box 252-21, 
      Hobart, TAS 7001, Australia
}

\date{16 October 1998}

\pagerange{\pageref{000}--\pageref{000}}
\pubyear{1999}

\begin{document}

\maketitle

\label{firstpage}

\begin{abstract}

We carried out $I$, $R$, $V$ and $B$ photometric observations of the neutron
star X-ray binary RXTE~J2123$-$058 shortly after the end of the X-ray
outburst in mid-1998. We adopt the low mass binary model to interpret our
observations. After folding our data on the 0.24821-d orbital period, and
correcting for the steady brightness decline following the outburst, we
observe sinusoidal oscillations with hints of ellipsoidal modulations which
became progressively more evident. Our data also show that the decline in
brightness was faster in the $V$ band than in the $R$ and $I$ bands. This
suggests both the cooling of an irradiation-heated secondary star and the
fading of an accretion disc over the nights of our observations. 

\end{abstract}


\section{Overview}
 
RXTE~J2123$-$058 was discovered as an X-ray transient by the Rossi 
X-ray Timing Explorer satellite ({\it RXTE}) in late June 1998 
(Levine, Swank \& Smith 1998). Its X-ray flux in the 2 -- 12 keV band 
was about 100~mCrab in the five measurements by the All Sky Monitor 
(ASM) on board {\it RXTE} on June 27 -- 28,  and was 65, 67 and 
56~mCrab at 2 -- 10~keV in the observations by the Proportional Counter 
Array (PCA) on June 27.09, 29.04 and 29.70 respectively 
(Takeshima \& Strohmayer 1998). Two X-ray bursts were detected on 
June 27.98 and 29.70. A preliminary analysis suggested that they were 
Type I bursts (Takeshima \& Strohmayer 1998).   

The optical counterpart of RXTE~J2123$-$058 was identified at the location
R.A. = $21^h 23^m 14.54^s$ and dec = $-5^{\circ}47'52.9''$ (equinox J2000;
Tomsick et al.\ 1998a).  Prior to the outburst it was barely visible as a
faint star in a digitised UK Schmidt plate.  During the outburst on June
30.44~UT the star brightened up to $U = 16.40$, $B = 17.28$, $V = 17.30$, $R
= 17.24$ and $I = 17.22$~mag (with an uncertainty of 0.05~mag).  The
$R$-band light-curve obtained between July 2 and 12 (Casares et al.\ 1998)
showed triangular-shaped minima with a central depth of 0.65~mag. Based on
this observation, a period of either 0.993 or 0.4965 d was proposed. The
$V$-band observations on June 30 -- July 4 and July 15 and 16 (Tomsick et
al.\ 1998b) showed  quasi-sinusoidal oscillations, with an amplitude 
of 0.9~mag and a best-fitting period of $5.957 \pm 0.003$~h ($0.24821 \pm
0.0001$~d).  The latter value is one-half of the 0.4965-d period that 
was previously proposed, but is consistent with the $5.9567 \pm
0.0033$~h period derived from
the $V$-band data obtained by Ilovaisky \& Chevalier (1998) from
independent observations on July 3 -- 18. The mean light-curve had a flat
top lasting a quarter of the orbital cycle and a broad triangular minimum
which spanned the rest of the cycle. At maximum light, the $V$-band
brightness was 16.8~mag. The ephemeris for the time of minimum brightness
was HJD~$2451009.888(3) + 0.2482(1)\times N$. Optical bursts of amplitude
$\sim 0.3$~mag were detected on June 30 and July 1 (Tomsick et al.\ 1998b). 

The X-ray brightness of RXTE~J2123$-$058 declined steadily after the
outburst on June 29 (see Fig. 1; quick-look results provided by the {\it
RXTE} team). By late August the one-day average count rates had dropped to
the pre-outburst level. The optical brightness also declined, at a rate of
$0.1\ {\rm mag\,d^{-1}}$ (Zurita \& Casares 1998). The average $R$
brightness of the source had dropped to 19.1~mag on August 16, but the
orbital modulation had increased from the previous value of 0.9~mag to
1.4~mag. Two narrow dips of 0.2~mag were observed at phases 0.0 and 0.5. From
these data, Zurita, Casares \& Hynes (1998) derived a revised ephemeris:
HJD~$2451042.639(5) + 0.24821(3) \times N$. On August 26 and 27, the $R$
brightness was $21.50\pm 0.06$~mag (Zurita \& Casares 1998) and the
light-curve began to exhibit a secondary minimum, i.e.\ an ellipsoidal
modulation due to the tidal distortion and the uneven surface brightness
distribution of the companion star. It is believed that the system had
returned to its quiescent state by then.  

We have carried out $V$-, $B$-, $R$- and $I$-band photometric 
observations of the X-ray transient RXTE~J2123$-$058 from 1998 August 
14 to 26. During our observations, the X-ray flux was in the final 
stage of decline and had almost reached the pre-outburst level. 
The system was in transition from an outburst to a quiescent state. 
Based on the currently available information, we propose a low-mass 
halo neutron-star binary model for the system, and attempt to 
interpret our photometric data in the light of the 
proposed model. 

\begin{figure}
\begin{center}
\epsfxsize=9.0cm 
\epsfbox{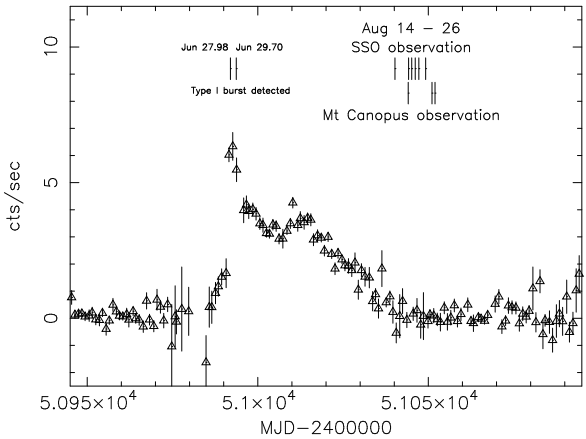} 
\end{center}
\caption{Dates of our photometric observations of RXTE~J2123$-$058, 
  in reference to its {\it RXTE}/ASM (2 -- 10~keV) X-ray light-curve 
  during the outburst. The dates where the two Type-I X-ray 
  bursts were detected are also marked. }
\end{figure} 
 
\section{Observation}    

The observations on August 14, 18, 19, 20, 21 and 23 were carried out 
with the Australian National University (ANU) 40-in telescope at the 
Siding Spring Observatory (SSO). We used a TEK $2048 \times 2048$ 
CCD mounted on the telescope. We took series of 300-s, 400-s and 600-s 
exposures in $V$, $R$ and $I$, and a few exposures in $B$ for 
colour calibration. The observing conditions were photometric during 
most of the runs. When the conditions were not photometric, the 
magnitude of RXTE~J2123$-$058 was determined from a comparison with 
a calibration star in the same field. Standard fields from the 
Landolt (1992) catalogue were observed on each night. 

The observations on August 18, 25 and 26 were carried out with the 
University of Tasmania (UTas) 1-m telescope at Mount Canopus. During 
the observations, the seeing was about 1.2 arcsec, and there were 
light clouds. We used an SBIG ST-6 $375\times 242$-pixel chip mounted 
at the Cassegrain focus. The exposure times were all 600~s.

\begin{figure}
\begin{center}
\epsfxsize=9.0cm 
\epsfbox{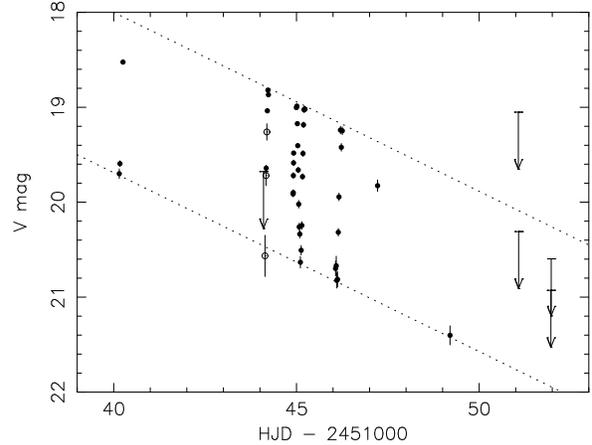} 
\end{center}
\caption{The decline in the $V$-band brightness during our 
  observations. The data obtained by the ANU 40-in telescope at Siding 
  Spring are represented by the filled circles, and the data obtained 
  by the UTas 1-m telescope at Mount Canopus by open circles. When 
  RXTE~J2123$-$058 was not visible, an upper limit was set by the 
  brightness of the faintest detectable star in the field. The two
  dotted lines 
  are the estimated cycle-to-cycle upper and lower limits to the 
  $V$-band magnitude for the data obtained by the ANU 40-in telescope. 
  The decline in $V$ is approximately linear between August 18 and 20. 
  The slope of the lines is 0.20~${\rm mag\,d^{-1}}$. The available data 
  are not sufficient to guarantee that the linear behaviour extends 
  beyond that time span. }
\end{figure}

\section{Photometric Light Curves}

The $V$-, $R$- and $I$-band data all show a large cycle-to-cycle modulation.
The $V$-band light curve obtained at SSO between August 18 and 20 is shown
in Fig.~2.  The minimum values measured within each observed cycle are
bounded by a straight line with a slope of 0.20~${\rm mag\,d^{-1}}$. The
corresponding maxima are also bounded by a straight line with the same
slope.   This suggests a (quasi-) linear decline in the $V$-brightness during
these three nights. However, the available data are not sufficient to
ascertain whether the linear behaviour extends to the whole time span of our
observations (August 14 to 23). 

In order to deconvolve the effect of the brightness decline  from the
orbital modulation, we apply a fit with sinusoidal and linear components to
the light curves.  Each data-point is then corrected to remove the effects
of the linear decline. We consider only the observations on August 19 -- 20
where the sampling is sufficient and the decline is presumably closest to
linear.  The best-fitting linear corrections in each band are $0.200 \pm
0.001\,{\rm mag\,d^{-1}}$ for $V$, $0.085 \pm 0.002\,{\rm mag\,d^{-1}}$ for
$R$ and $0.020 \pm 0.010\,{\rm mag\,d^{-1}}$ for $I$ (top and middle panels in
Fig.~3, 4 and 5). The corrected phase-dependent $V$-, $R$-, and $I$-band
light-curves (folded with a period of 0.24821~d) are shown in the bottom
panels of Fig.~3, 4 and 5. 

We also observed the system in the $B$ band briefly on August 14 and 18, 
(only 6 images taken). The $B$ brightness (not shown) varied between 
18.5 and 20.0~mag during our observations.

The photometric data in different bands were not taken simultaneously, 
therefore we cannot calculate the colour excess without interpolations;  
however, the rapid decline in brightness and the sparse sampling 
during our observations do not permit a reliable interpolation.   

\begin{figure}
\begin{center}
\epsfxsize=8cm 
\epsfbox{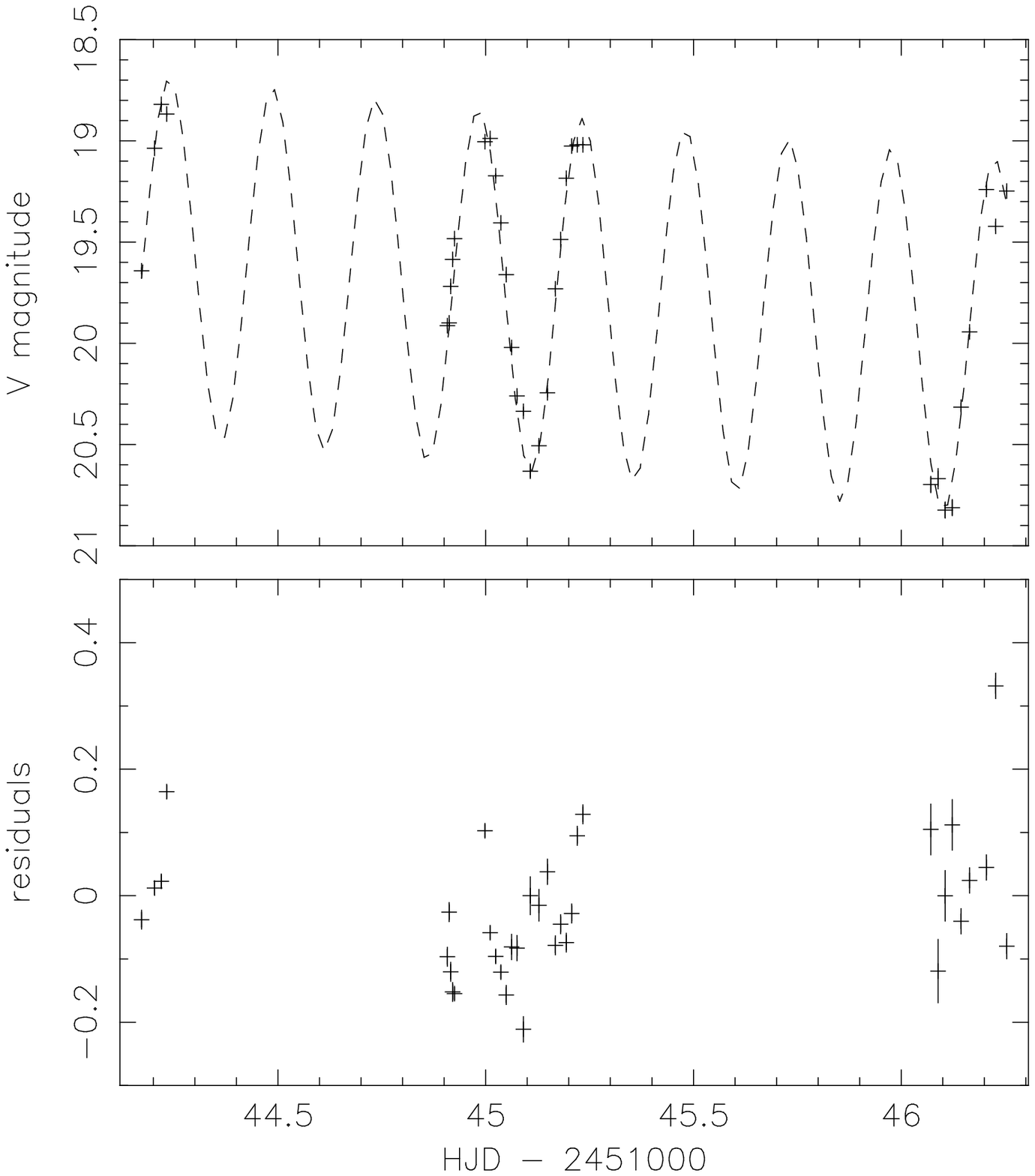} 
\epsfxsize=8cm 
\epsfbox{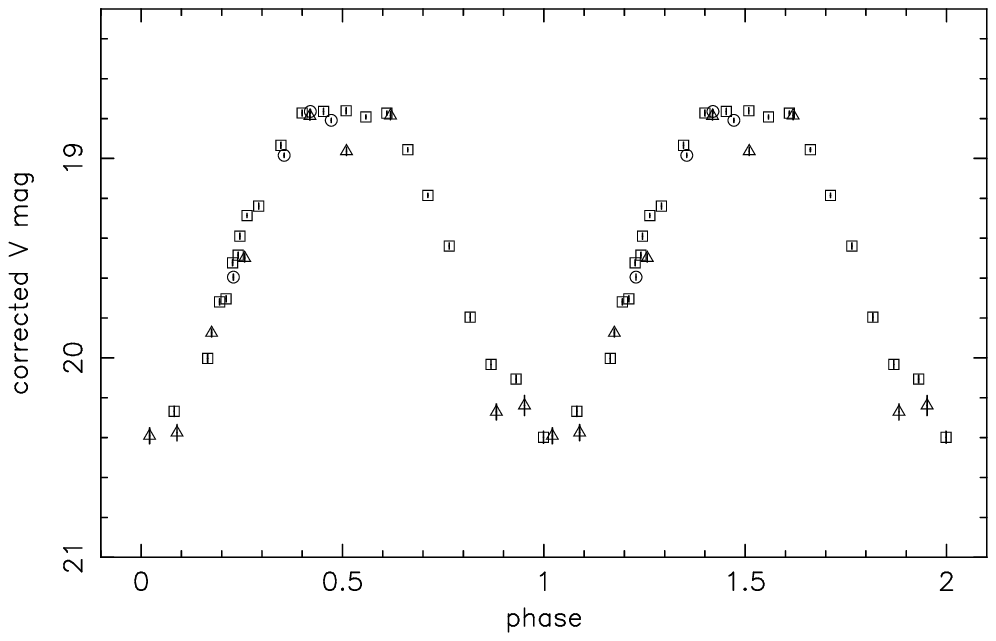} 
\end{center}
\caption{The best sinusoidal fit to the $V$-band light-curve from the August
  18 -- 20 observations (top),  the residuals of the fit (middle) and the
  folded $V$-band light-curve (bottom). The rate of decline in the $V$
  brightness  deduced from the fit is $0.200 \pm 0.001\,{\rm mag\,d^{-1}}$,
  and the amplitude of the brightness modulations is 0.884~mag. The period is
  kept fixed at 0.24821~d in our fit.  Dates for the observations are August
  18 (circle), 19 (square) and 20 (triangle).}
\end{figure}   

\begin{figure}
\begin{center}
\epsfxsize=8cm 
\epsfbox{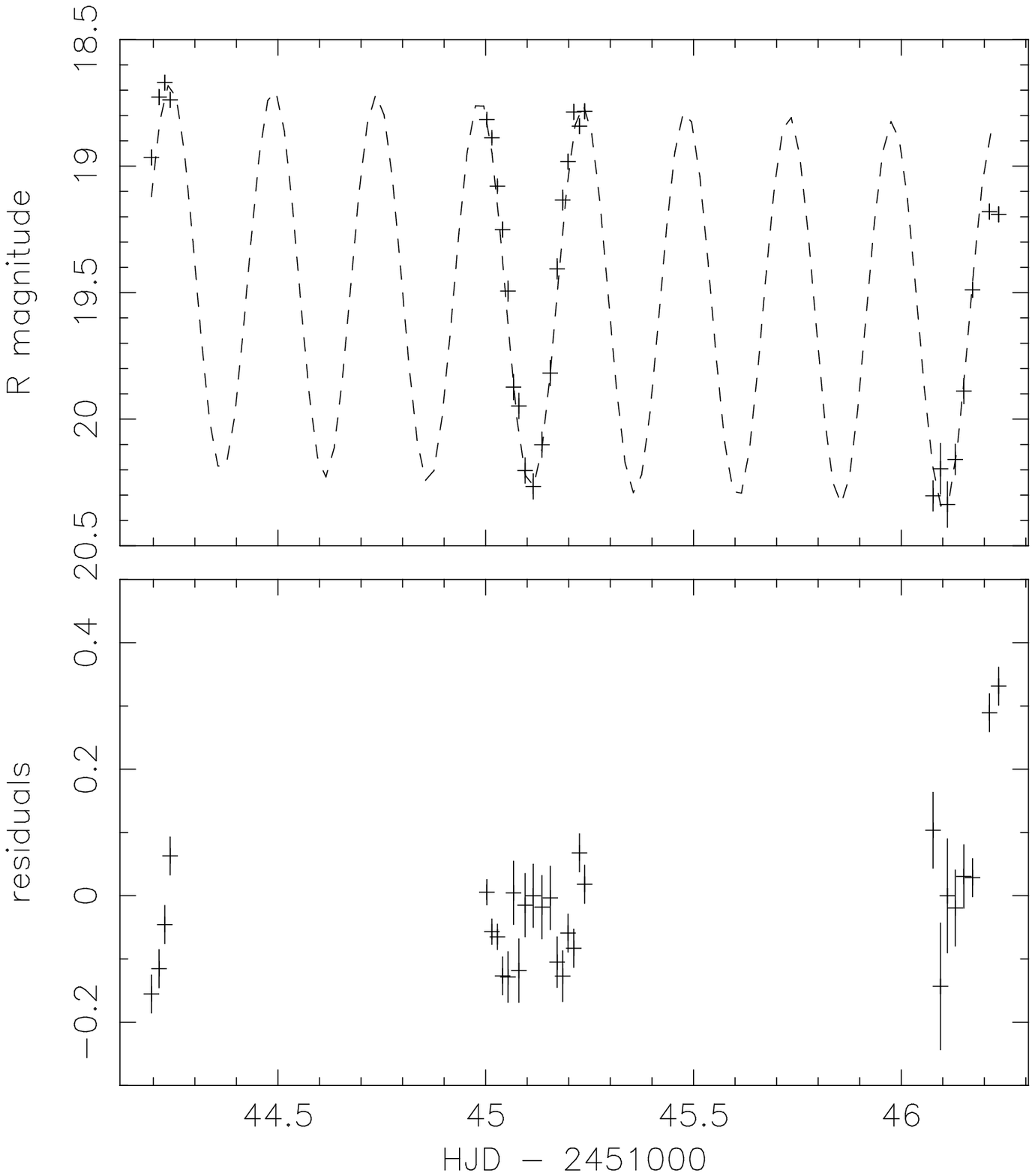} 
\epsfxsize=8cm 
\epsfbox{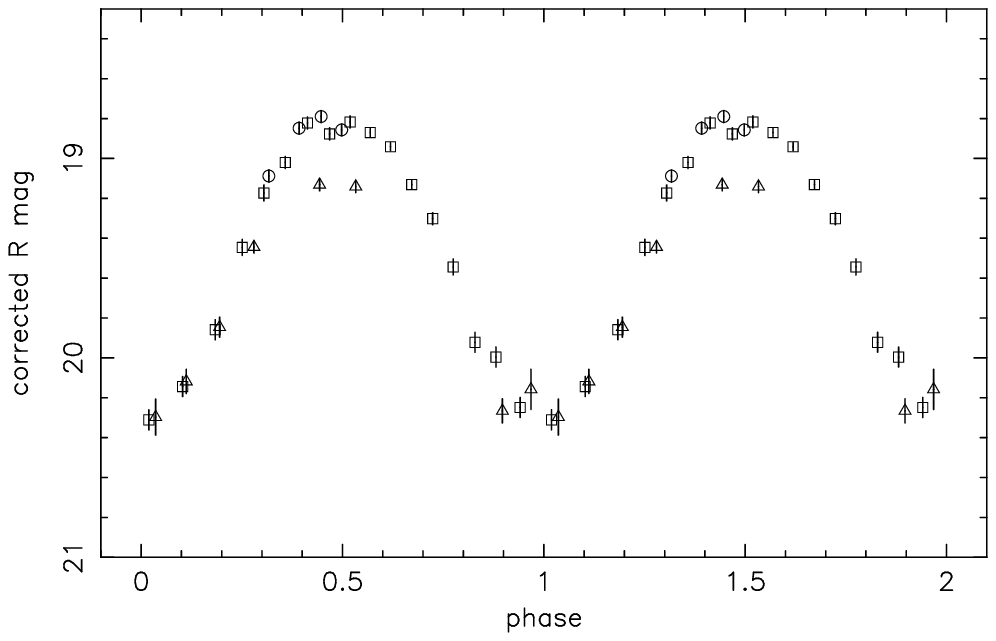} 
\end{center}
\caption{The best sinusoidal fit to the $R$-band light-curve from  the
  August 18 -- 20 observations (top),  the residuals of the fit (middle) and
  the folded $R$-band light-curve (bottom). The rate of decline in the $R$
  brightness  deduced from the fit is $0.085 \pm 0.002\,{\rm mag\,d^{-1}}$,
  and the amplitude of the brightness modulation is 0.761~mag. The period is
  kept fixed at 0.24821~d in our fit.  Dates for the observations are August
  18 (circle), 19 (square) and 20 (triangle).}
\end{figure}    

\begin{figure}
\begin{center}
\epsfxsize=8cm 
\epsfbox{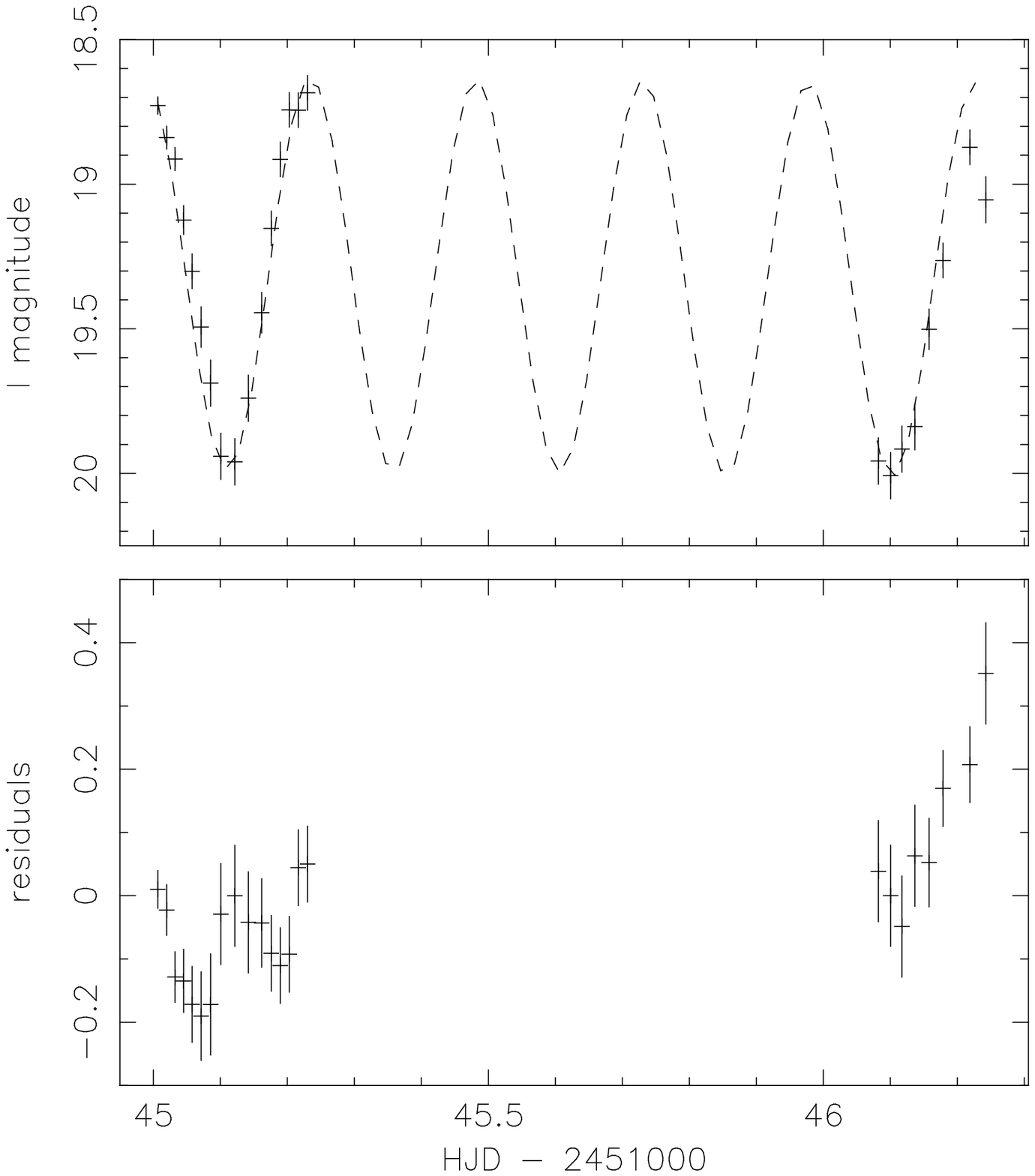} 
\epsfxsize=8cm 
\epsfbox{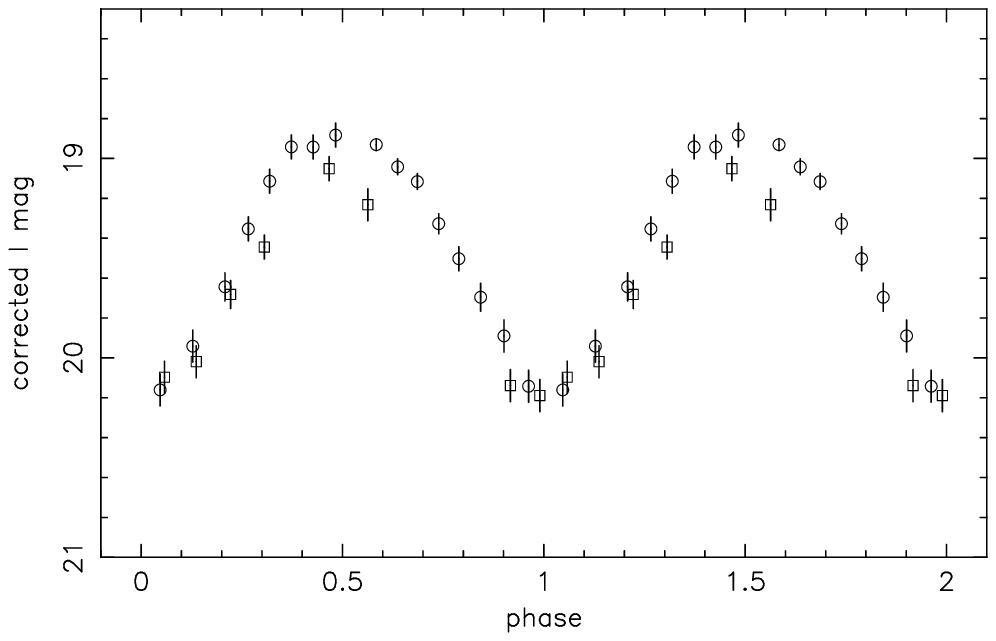} 
\end{center}
\caption{ The best sinusoidal fit to the $I$-band light-curve from the
  August 19 -- 20 observations (top),  the residuals of the fit (middle) and
  the folded $R$-band light-curve (bottom). The rate of decline in the $I$
  brightness  deduced from the fit is $0.020 \pm 0.010\,{\rm mag\,d^{-1}}$,
  and the amplitude of the brightness modulations is 0.681~mag. The period is
  fixed to be 0.24821~d in the fit. Dates for the observations are August 19
  (circle) and 20 (square). }
\end{figure}   

\begin{figure}
\begin{center}
\epsfxsize=9.0cm 
\epsfbox{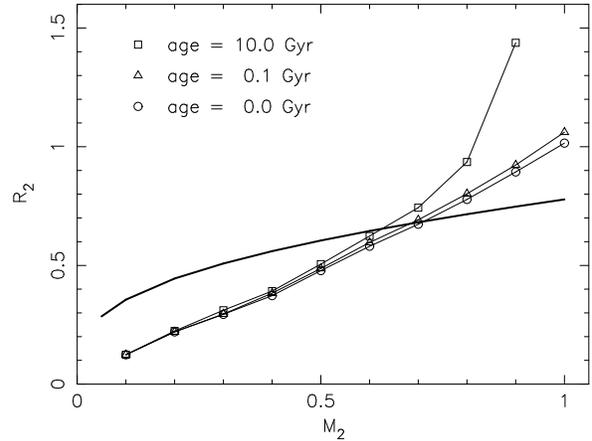} 
\end{center}
\caption{The Roche-lobe radius (thick line) of the companion star 
  in a neutron star binary with a circular orbit and an orbital period 
  of 0.24821~d, and the radius of the companion star for various ages. 
  The Roche-lobe radius is calculated using the Eggleton (1983) formula 
  with the assumption that the mass of the neutron star is 1.4~M$_\odot$. 
  The radius of the star is derived from the luminosity-temperature 
  relation in the evolutionary models given in Baraffe et\,al. (1998). 
  Units are solar masses and solar radii on the x- and y-axis respectively. }
\end{figure} 

\section{The Nature of the System} 

The detection of Type-I X-ray bursts implies that the compact star in
RXTE~J2123$-$058 is a neutron star. The distance determined by assuming an
Eddington flux during the X-ray burst is about 14~kpc (Homan et\,al.\ 1999).
However, the peak flux during the burst might not be at the Eddington limit.
More often X-ray bursts are weaker, i.e. the peak luminosities are
sub-Eddington (see e.g.\ Lewin, van Paradijs \& Taam 1995). When the
persistent flux is high (which indicates a high accretion rate), the
hydrogen-helium rich matter on the surface of the neutron star might be
heated to the ignition temperature for the thermonuclear reaction such that
``premature'' bursts may occur.  If we conservatively estimate that the
peak luminosity during the peak of the X-ray burst is only 1/10 of the
Eddington luminosity (there is for example a factor of about 6 to 7
between the maximum and minimum peak burst luminosities in 4U/MXB
1636-53, see Fujimoto et\,al.\ 1988),
then the corresponding distance will be reduced by a factor of 3. This will
lower the estimated distance of the source to about 5~kpc. 

The 0.24821-d period suggests that the system is a low-mass X-ray binary
(see the catalogue of X-ray binaries compiled by van Paradijs 1995).  In a
low-mass system, the companion star must fill its Roche lobe to allow mass
transfer to occur. We show in Fig.~6 the radius $R_{\rm h}$ of the companion
star's Roche lobe (Eggleton 1983) as a function of its mass $M_2$ for a
circular binary orbit and a 1.4~M$_\odot$ primary.  In the same diagram we
also show the radii of stars in the mass range from 0.1 to 1.0~M$_\odot$
derived from the evolutionary models given in Baraffe et\,al.\ (1998), at
ages 0, 0.1 and 10~Gyr. The requirement that the companion star's radius
$R_2$ is approximately equal to its Roche-lobe radius $R_{\rm h}$ implies
$R_2 \la 0.7 R_\odot$ and consequently restricts the mass of the companion
star to be below 0.7~M$_\odot$ if it is a hydrogen main-sequence star.
Although a different mass limit can be obtained for helium stars, the
detection of hydrogen absorption lines (Tomsick et\,al.\ 1998a) makes it
unlikely that the companion star in RXTE~J2123$-$058 be a helium star. The
actual mass is probably slightly lower than 0.7~M$_\odot$, as the star might
have evolved off the main-sequence. Moreover, the pre-outburst X-ray
luminosity was weak, indicating that the star underfilled its Roche lobe.
A low-mass companion star, a distance of $\sim 10$~kpc and galactic
coordinates of $l = 46^{\circ} 28' 58.4''$ and $b = -36^{\circ} 11' 57.3''$
imply that RXTE~J2123$-$058 is an old low-mass X-ray binary in the galactic
halo.    

\section {Companion Star} 

For a 0.7~M$_\odot$ star with an age of about $10^8 - 10^9$~yr, the absolute 
magnitudes in the $V$, $R$ and $I$ bands are $M_{_V} \approx 6.8$,  
$M_{_R} \approx 6.2$ and $M_{_I} \approx 5.7$ (assuming a metal abundance 
${\rm [M/H]} = -0.5$ and Y = 0.25) (Baraffe et\,al.\ 1998). At a distance 
of 15~kpc, the corresponding apparent magnitudes (neglecting extinction) 
are $m_{_V} \approx 22.7$, $m_{_R} \approx 22.1$ and 
$m_{_I} \approx 21.6$. If we take a distance of 5~kpc, then the apparent 
magnitudes are $m_{_V} \approx 20.3$, $m_{_R} \approx 19.7$ 
and $m_{_I} \approx 19.2$.  As RXTE~J2123$-$058 is probably a halo 
source, the extinction of its counterpart can be estimated from the 
reddening derived from the H{\small{ I}} measurement and galaxy counts. 
From the reddening map in Burstein \& Heiles (1982) we obtain 
a colour excess $E(B-V) \approx 0.054$ for a halo source at 
$l = 46^{\circ} 28' 58.4''$ and $b = -36^{\circ} 11' 57.3''$. The 
corresponding extinctions are $A_V \approx 0.17$, $A_R \approx 0.11$ and 
$A_I \approx 0.09$. Hence, the extinction-corrected magnitudes of a 
0.7~M$_\odot$ star at a distance of 15~kpc are $m_{_V} \approx 22.9$, 
$m_{_R} \approx 22.2$ and $m_{_I} \approx 21.7$.  

From the PCA count rates and the distance estimate ($5 - 15$~kpc), we obtain
an estimate of $\sim 1 \times 10^{37}~{\rm erg\,s^{-1}}$ for the X-ray
luminosity during MJD~$\approx 2450995 - 2451105$.  A fraction $(\pi
R_2^2)/(4 \pi a^2)$ of the X-rays would intercept the companion star, where
$R_2$ is the radius of the companion star and $a$ the orbital separation.
If we substitute the values of $R_2$ and $M_2$ calculated in \S4, this
corresponds to a fraction of $\sim 3$~per~cent. Therefore, energy will be
deposited into the companion star's atmosphere at a rate of $\sim 5 \times
10^{35}~{\rm erg\,s^{-1}}$.  An unheated 0.7~M$_\odot$ main-sequence star
has a bolometric luminosity of about $1 \times 10^{33}~{\rm erg\,s^{-1}}$
and an effective surface temperature of about 4800~K. The intrinsic
luminosity of the companion star is therefore much lower than the power of
the intercepted X-rays. We suppose that a quasi-equilibrium state is set up
at the heated surface of the companion star such that the rate of energy
radiated away is the same as the rate of energy deposited. The effective
surface temperature of the irradiatively-heated atmosphere of the companion
star could then reach 20000~K.  

During our observations, the system was in the process of returning to its
quiescent state. On August 23 the ASM count rate  had already dropped to the
pre-outburst level (Fig.~1), which may be considered consistent with zero.
Although the X-ray activity seemed to have ceased, the accretion disc had
not completely dissipated, and the companion star's atmosphere had not
completely cooled down. Our photometric data obtained on August 19 and 20
show that the minimum brightness in the $V$-, $R$- and $I$-band light-curves
was $V \approx 20.8$, $R \approx 20.5$ and $I \approx 20.5$. The $V$-band
brightness continued to drop and reached 21.4~mag on August 23. The star was
finally not visible in the $B$-, $R$- and $V$-band data that we obtained on
August 25 and 26 (cf.\ the $R$-band brightness of 21.5~mag on August 26 and
27, Zurita \& Casares 1998, which would imply a distance of $\sim 11$ kpc if
it is taken as the quiescent brightness of the companion star).

The optical brightness of RXTE~J2123$-$058 was observed to decline at a rate
of 0.1~${\rm mag\,d^{-1}}$ in early August (Zurita, Casares \& Hynes 1998).
We observed a decline rate $0.200 \pm 0.001\,{\rm mag\,d^{-1}}$ in the
$V$-band brightness between August 18 and 20.  A decline is also seen in our
$R$- and $I$-band data but at less rapid rates of $0.085 \pm 0.002\,{\rm
mag\,d^{-1}}$  and $0.020 \pm 0.010\,{\rm mag\,d^{-1}}$ respectively. A
linear decline in brightness is equivalent to an exponential decline in
luminosity. The faster decline in the $V$ magnitude may indicate the fading
or dissipation of the accretion disc (which was bluer than the companion
star) after mass transfer ceased, in addition to the cooling of the
atmosphere of the companion star. The photometric data also show hints of an
ellipsoidal modulation around phase 0.5, which became more evident in our
later observations. As we show in Fig.~3, the $V$-band light-curve deviates
from a sinusoidal-like curve by developing first a flat top around phase 0.5
on August 18, and then a local minimum on August 20. We attribute the
gradual development of the secondary minimum at phase 0.5 to the fading of
the accretion disk, which reveals the ellipsoidal modulation of the
companion star.

\section{Summary}

We carried out a synergetic study of the X-ray transient RXTE~J2123$-$058.
The photometric observations were carried out from 1998 August 14 to 26,
when the system was in transition from the outburst to the quiescent state.
High-quality data were obtained in the $V$, $R$ and $I$ bands, which show
approximately sinusoidal oscillations with a linear brightness decline when
folded on a period of 0.24821~d. The rate of decline measured from the $V$
brightness during our observations was $0.200 \pm 0.001\,{\rm mag\,d^{-1}}$,
more rapid than that the rate of $0.085 \pm 0.002\,{\rm mag\,d^{-1}}$ for
the $R$ brightness and of $0.020 \pm 0.010\,{\rm mag\,d^{-1}}$ for the $I$
brightness.  The folded light-curves deviate significantly from a sinusoidal
curve around phase 0.5, showing hints of ellipsoidal modulations, which tend
to become progressively more evident.  This suggests the presence of a
fading accretion disc as well as an irradiatively-heated companion star.  

\section*{Acknowledgements} 

We thank Allyn Tennant, Helen Johnston, Mike Bessell and Peter Wood 
for discussions, and Don Melrose and John Greenhill for their comments 
on the manuscript. We also thank John Greenhill and Jorge Casares for 
their sharing some of the data obtained by the UTas 1-m telescope. KW 
acknowledges the support from the ARC through an Australian Research 
Fellowship. This work is partially supported by the URG, University of 
Sydney.

\end{document}